\documentclass[a4paper,UKenglish,cleveref, autoref, thm-restate]{lipics-v2021}

\nolinenumbers

\hideLIPIcs  

\title{Small Scale Reflection for the Working Lean User} 

\author{Vladimir Gladshtein}{
School of Computing, National University of Singapore, Singapore
\and
\url{https://volodeyka.github.io/}}{vgladsht@comp.nus.edu.sg}{https://orcid.org/0000-0001-9233-3133}{}

\author{George P\^{i}rlea}{
School of Computing, National University of Singapore, Singapore
\and
\url{https://pirlea.net/}}{gpirlea@comp.nus.edu.sg}{https://orcid.org/0009-0008-5378-2815}{}

\author{Ilya Sergey}{
School of Computing, National University of Singapore, Singapore
\and
\url{https://ilyasergey.net/}}{ilya@nus.edu.sg}{https://orcid.org/0000-0003-4250-5392}{}

\authorrunning{V. Gladshtein, G. P\^{i}rlea, and I. Sergey} 


\usepackage{lipsum}
\usepackage{graphicx}
\usepackage{mathpartir}
\usepackage{xspace}
\usepackage{bbding}
\usepackage{pifont}
\usepackage{syntax}
\usepackage{float}
\usepackage{proof}
\usepackage{stmaryrd}
\usepackage{calc}
\usepackage[most]{tcolorbox}
\usepackage{tikz-cd}
\usepackage{mathtools}
\usepackage{thm-restate}
\usepackage{cite}
\usepackage{booktabs}
\usepackage[inline, shortlabels]{enumitem}
\usepackage{listings}
\usepackage{amssymb}
\usepackage{wrapfig}
\usepackage[figurename=Fig.,labelsep=period]{caption}
\usepackage{amsmath}

\usepackage{tikz}
\makeatletter
\newenvironment{btHighlight}[1][]
{\begingroup\tikzset{bt@Highlight@par/.style={#1}}\begin{lrbox}{\@tempboxa}}
{\end{lrbox}\bt@HL@box[bt@Highlight@par]{\@tempboxa}\endgroup}

\newcommand\btHL[1][]{%
  \begin{btHighlight}[#1]\bgroup\aftergroup\bt@HL@endenv%
}
\def\bt@HL@endenv{%
  \end{btHighlight}%
  \egroup
}
\newcommand{\bt@HL@box}[2][]{%
  \tikz[#1]{%
    \pgfpathrectangle{\pgfpoint{1pt}{0pt}}{\pgfpoint{\wd #2}{\ht #2}}%
    \pgfusepath{use as bounding box}%
    \node[anchor=base west, fill=orange!30,outer sep=0pt,inner xsep=1pt, inner ysep=0pt, rounded corners=3pt, minimum height=\ht\strutbox+1pt,#1]{\raisebox{1pt}{\strut}\strut\usebox{#2}};
  }%
}
\makeatother

\usepackage{color}
\definecolor{keywordcolor}{rgb}{0.7, 0.1, 0.1}   
\definecolor{tacticcolor}{rgb}{0.0, 0.1, 0.6}    
\definecolor{commentcolor}{rgb}{0.4, 0.4, 0.4}   
\definecolor{symbolcolor}{rgb}{0.0, 0.1, 0.6}    
\definecolor{sortcolor}{rgb}{0.1, 0.5, 0.1}      
\definecolor{attributecolor}{rgb}{0.7, 0.1, 0.1} 

\makeatletter 
\def\arcr{\@arraycr}
\makeatother

\definecolor{shadecolor}{gray}{1.00}
\definecolor{darkgray}{gray}{0.30}
\definecolor{violet}{rgb}{0.56, 0.0, 1.0}
\definecolor{forestgreen}{rgb}{0.13, 0.55, 0.13}
\definecolor{mygray}{rgb}{0.5,0.5,0.5}

\definecolor[named]{ACMBlue}{cmyk}{1,0.1,0,0.1}
\definecolor[named]{ACMYellow}{cmyk}{0,0.16,1,0}
\definecolor[named]{ACMOrange}{cmyk}{0,0.42,1,0.01}
\definecolor[named]{ACMRed}{cmyk}{0,0.90,0.86,0}
\definecolor[named]{ACMLightBlue}{cmyk}{0.49,0.01,0,0}
\definecolor[named]{ACMGreen}{cmyk}{0.20,0,1,0.19}
\definecolor[named]{ACMPurple}{cmyk}{0.55,1,0,0.15}
\definecolor[named]{ACMDarkBlue}{cmyk}{1,0.58,0,0.21}
\definecolor[named]{PaleGreen}{RGB}{196, 255, 231}
\definecolor[named]{PaleOrange}{RGB}{255, 213, 169}
\definecolor{intnull}{RGB}{213,229,255}

\definecolor{shadecolor}{gray}{1.00}
\definecolor{ddarkgray}{gray}{0.85}
\definecolor{darkgray}{gray}{0.30}
\definecolor{light-gray}{gray}{0.91}

\newcommand{\graybox}[1]{\colorbox{ddarkgray}{#1}}

\newcommand{\redbox}[1]{\colorbox{red!30}{#1}}
\newcommand{\orangebox}[1]{\colorbox{orange!30}{#1}}
\newcommand{\gbm}[1]{\graybox{${#1}$}}

\definecolor{ipurple}{HTML}{E5D8F3}
\definecolor{igreen}{HTML}{C2F4CF}
\definecolor{ired}{HTML}{AE0428}
\definecolor{iblue}{HTML}{3E74D1}
\definecolor{ilgray}{HTML}{E3E3E3}
\definecolor{idgray}{HTML}{747474}

\newcommand{\etc}{\emph{etc}\xspace}
\newcommand{\ie}{\emph{i.e.}\xspace}

\newcommand{\eg}{\emph{e.g.}\xspace}

\newcommand{\aka}{\textit{a.k.a.}\xspace}
\newcommand{\cf}{\textit{cf.}\xspace}

\newcommand{\wrt}{\emph{w.r.t.}\xspace}

\newcommand{\tname}[1]{\textsf{#1}\xspace}

\newcommand{\ssr}{\tname{SSReflect}}
\newcommand{\lssr}{\tname{LeanSSR}}

\definecolor{pblue}{rgb}{0.13,0.13,1}
\definecolor{pgreen}{rgb}{0,0.5,0}
\definecolor{pred}{rgb}{0.9,0,0}
\definecolor{pgrey}{rgb}{0.46,0.45,0.48}

\definecolor{ckeyword}{HTML}{7F0055}
\definecolor{ccomment}{HTML}{3F7F5F}
\definecolor{cnumber}{HTML}{2A0099}

\newcommand{\code}[1]{{\lstinline[basicstyle=\ttfamily]{#1}}\xspace}
\newcommand{\fcode}[1]{\lstinline[basicstyle=\ttfamily\small]{#1}\xspace}

\makeatletter
\protected\def\ccell#1#{%
  \kern-\fboxsep
  \@ccell{#1}%
}
\def\@ccell#1#2#3{%
  \colorbox#1{#2}{#3}%
  \kern-\fboxsep
}
\makeatother

\newcommand{\hide}[1]{}

\makeatletter
\providecommand*{\cupdot}{%
  \mathbin{%
    \mathpalette\@cupdot{}%
  }%
}
\newcommand*{\@cupdot}[2]{%
  \ooalign{%
    $\m@th#1\cup$\cr
    \hidewidth$\m@th#1\cdot$\hidewidth
  }%
}
\makeatother

\usetikzlibrary{decorations.pathmorphing}

\tikzset{snake it/.style={decorate, decoration=snake}}

\def\signed #1{{\leavevmode\unskip\nobreak\hfil\penalty50\hskip1em
  \hbox{}\nobreak\hfill #1%
  \parfillskip=0pt \finalhyphendemerits=0 \endgraf}}

\newsavebox\mybox

\definecolor{grown}{rgb}{0.6, 0, 0.6}
\definecolor{blue}{rgb}{0, 0, 0.9}

\providecommand{\rule}[3]{
  \begin{prooftree}
    \AXC{\visible<2->{#1}}
    \RightLabel{\visible<2->{#2}}
    \UIC{#3}
  \end{prooftree}
}

\hypersetup{colorlinks,
  linkcolor=ACMDarkBlue,
  citecolor=ACMPurple,
  urlcolor=ACMDarkBlue,
  filecolor=ACMDarkBlue}

\floatname{algorithm}{Algorithm}

\setlist[itemize]{leftmargin=*}
\setlist[enumerate]{leftmargin=*}

\begin{document}

\maketitle

\begin{abstract}
We present the design and implementation of the Small Scale Reflection proof methodology and tactic language (\aka SSR) for the Lean 4 proof assistant.
Like its Coq predecessor \ssr, 
%
%
our Lean 4 implementation, dubbed \lssr, provides powerful rewriting principles and means for effective management of hypotheses in the proof context.
Unlike \ssr for Coq, \lssr does not require explicit switching between the \emph{logical} and \emph{symbolic} representation of a goal, allowing for even more concise proof scripts that seamlessly combine deduction steps with proofs by computation.

In this paper, we first provide a gentle introduction to the principles of structuring mechanised proofs using \lssr.
Next, we show how the native support for metaprogramming in Lean 4 makes it possible to develop \lssr entirely within the proof assistant, greatly improving the overall experience of both tactic implementers and proof engineers.
Finally, we demonstrate the utility of \lssr by conducting two case studies: (a)~porting a collection of Coq lemmas about sequences from the widely used Mathematical Components library and (b)~reimplementing proofs in the finite set library of Lean's mathlib4.
Both case studies show significant reduction in proof sizes.


\end{abstract}

\section{Introduction}
\label{sec:intro}

\emph{Small Scale Reflection} (SSR) is a methodology for structuring
deductive machine-assisted proofs that promotes the pervasive use of
\emph{computable symbolic} representations of data properties, in
addition to their more conventional logical definitions in the form of
inductive relations.
Small scale reflection emerged from the prominent effort to
mechanise the proof of the Four Colour Theorem in the Coq proof
assistant~\cite{GonthierFCT}, in which the large number of cases to
be discharged posed a significant scalability challenge for a traditional
proof style based on tactics that operate directly with the logical
representation of a goal.
Support for small scale reflection in Coq has later been implemented
in the form of the \ssr plugin, which provides a concise tactic
language, and its associated library of lemmas~\cite{Gonthier-al:TR},
becoming an indispensable tool for the working Coq user.~\ssr has been
employed in many projects, including mechanisations of the foundations
of group theory~\cite{GonthierAABCGRMOBPRSTT13}, measure
theory~\cite{AffeldtC23}, information theory~\cite{AffeldtHS14}, 3D
geometry~\cite{AffeldtC17},
programming language semantics~\cite{WeirichVAE17}, as well as proving the
correctness of heap-manipulating, concurrent and distributed
programs~\cite{Nanevski-al:POPL10,SergeyNB15,SergeyWT18,PirleaS18},
and probabilistic data structures~\cite{GopinathanS20}.
Two \ssr tutorials~\cite{Maboubi-Tassi:MathComp,Sergey:PnP} are
currently recommended amongst the basic learning materials for
Coq~\cite{coqdoc}.

Lean~4 is a relatively new proof assistant and dependently typed
programming language~\cite{lean}.\footnote{In the rest of the paper, we
  will refer to the latest version 4 of the framework simply as Lean.}
Like Coq, Lean is based on the Calculus of Constructions with
inductive types and is geared towards interactive proofs, coming with
extensive support for metaprogramming aimed at simplifying custom
proof automation and code generation.
Unlike Coq, Lean assumes axioms of classical logic, such as the law of
excluded middle.
Very much in the spirit of the SSR philosophy of \emph{reflecting}
proofs about decidable propositions into Boolean-returning
\emph{computations},
Lean encourages the use of such propositions (implemented as an
instance of the \code{Decidable} type class) \emph{as if they were}
Boolean expressions,
going as far as providing a program-level \code{if-then-else} operator
that performs conditioning on an instance of a decidable proposition.
Given this similarity of approaches, it was only natural for us to try
to bring the familiar SSR tactic language to Lean, as an alternative
to Lean's descriptive but verbose idiomatic approach to proof
construction.
Our secondary motivation was to put Lean's metaprogramming to the
test, implementing SSR entirely within the proof assistant, in
contrast with \ssr, which is implemented as a Coq plugin in OCaml.

In this paper, we present \lssr, a proof scripting language that
faithfully replicates the \ssr experience in Lean, yet provides
\emph{substantially better} proof ergonomics, thanks to Lean's
distinctive features, improving on its Coq predecessor in the
following three aspects:

\begin{enumerate}[topsep=0pt]
\item\label{label:ui} \emph{Usability}: Compared to Coq/\ssr,
  which only shows the proof context between complete series of
  chained tactic applications, \lssr provides a more fine-grained
  access to the proof state, displaying its changes after executing
  each ``atomic'' step---a feature made possible by Lean's mechanism
  of proof state annotations.
\item\label{label:express} \emph{Expressivity}: Unlike \ssr for Coq,
  \lssr does not require the user to explicitly switch between
  representation of facts as logical propositions or as symbolic
  expressions to advance the proof.
  %
  %
  In particular, this automation makes it possible to unify
  simplifications via computation and via equality-based rewriting,
  resulting in very concise proof scripts.
\item\label{label:ext} \emph{Extensibility}: Since \lssr is
  \emph{lightweight}, \ie, it is implemented entirely in Lean using
  its metaprogramming facilities, it can be easily enhanced with
  additional tactics and proof automation machineries specific to a
  particular project that uses it.
\end{enumerate}

\noindent
In the rest of the paper, we showcase \lssr, substantiating the three
claims above.

\vspace{3pt}
\noindent
\emph{Contributions and Outline.~}
In this work, we make the following contributions:
%

\begin{itemize}[topsep=0pt]
\item An implementation of \lssr, an \ssr-style tactic language for
  Lean.\footnote{The snapshot of \lssr development accompanying this
    submission is available online~\cite{lssr}.}
\item A tutorial on using \lssr following a series of characteristic
  proof examples (\autoref{sec:overview}).
\item A detailed overview of Lean metaprogramming features that we
  consider essential for implementing \lssr and the effect of those
  features on the end-user experience (\autoref{sec:implementation}).
\item Two case studies demonstrating \lssr's utility for proof
  migration and evolution: (a)~porting a collection of Coq lemmas
  about finite sequences from the MathComp
  library~\cite{Maboubi-Tassi:MathComp} to Lean and (b)~reimplementing
  proofs from the finite set library of Lean's mathlib4 in \lssr
  (\autoref{sec:case}).
  While proof sizes reduce in both cases, we also comment on our
  overall better mechanisation experience, compared to refactoring
  original proofs in both Coq and Lean---thanks to the improved
  usability aspect highlighted above.
\end{itemize}

%


\section{\lssr in Action}
\label{sec:overview}


In this section we give a tutorial on the main elements of the \lssr
proof language. We do not expect the reader to be familiar with the
standard Lean tactics. This section will also be informative for the
readers familiar with \ssr for Coq due to the improvements \lssr makes
over \ssr in terms of usability (\autoref{sec:debug}) and expressivity
(\autoref{sec:refl}).

\vspace{-5pt}

\subsection{Managing the Context and the Goal in Backward Proofs}
\label{sec:move}

\noindent
As customary for interactive proof assistants based on higher-order
logic, Lean represents the proof state as a logical \emph{sequent}, as
depicted in \autoref{fig:goal}.

\begin{wrapfigure}[7]{r}{0.35\textwidth}
    \vspace{-15pt}
    \captionsetup{aboveskip=5pt}
    \begin{mathpar}
        \inferrule[]
        {
            \begin{array}{l}
            c_i : T_i \arcr
            \dots \arcr
            F_k : P_k \arcr
            \dots \arcr
            \end{array}
        }
        {
            \begin{array}{l}
            \forall x_l : T_l, \ \dots, 
            P_n \to \dots \to C 
            \end{array}
        }
    \end{mathpar}
    \vspace{-3mm}
    \caption{Proof state as a sequent}
    \label{fig:goal}
\end{wrapfigure}
The proof \emph{goal} is the logical statement below the horizontal
line; above the line is the \emph{local context} of the sequent, a set
of constants $c_i$ and facts (\ie, hypotheses) $F_k$. The goal
statement itself can be decomposed into~(a)~\emph{goal context}:
\emph{quantified variables} $x_l : T_l$ and \emph{assumptions} $P_n$,
and~(b)~the \emph{conclusion} statement $C$.
Both local and goal contexts capture the bound variables (term-level
or hypotheses) of the overall proof term to be constructed, with the
only difference being the explicit names of the variables in the local
context.
Lean tactics, such as \code{apply}, operate directly on the goal's
conclusion using variables from the local context referring to them by
their names, while tactics such as \code{intro} and \code{revert} move
variables/hypotheses between the local and the goal context.
In practice, such ``bookkeeping'' of variables and hypotheses
contributes most of the proof script burden; so, following \ssr,
\lssr provides tactics to streamline it.


Specifically, \lssr provides mechanisms to operate directly with the
goal context, treating the goal as a \emph{stack}, where the left-most
variable or assumption is the top of the stack, and the conclusion is
always at the bottom of the stack. 
By convention, most of \lssr tactics, operate with the top element of
the goal stack. 
As an example, \lssr defines \code{sapply}, a variant of the standard
Lean \code{apply} tactics (many \lssr tactics come with the prefix
\code{s*} to distinguish them from their standard Lean counterparts) to
simplify backward proofs~\cite[Chapter~3]{love}.
As \autoref{fig:sapply} shows, \code{sapply} applies the first element
on the goal stack (\ie, the hypothesis of type $\alpha$) to the rest
of the stack (\ie, the goal's conclusion $\alpha$).

\begin{figure}[b]
    \begin{subfigure}{0.26\textwidth}
        \begin{lstlisting}
example: α -> α := 
    by sapply
        \end{lstlisting}
        \caption{Using \fcode{sapply} tactic}
        \label{fig:sapply}
    \end{subfigure}
    \begin{subfigure}{0.30\textwidth}
        \begin{lstlisting}
example: @a -> @b -> @a :=
  by move=> hA ?
     move: hA; sapply\end{lstlisting}
    \caption{Basic \lssr proof script}
    \label{fig:inro-rev1}
    \end{subfigure}
    \begin{subfigure}{0.40\textwidth}
        \begin{lstlisting}
example: @a -> (@a -> @b) -> @b :=
    by move=> /[swap] hAiB
       sapply: hAiB\end{lstlisting}
    \caption{A proof using \fcode{/[swap]} intro pattern}
    \label{fig:inro-rev2}
    \end{subfigure}
    \caption{\lssr proof scripts using the \fcode{sapply} tactic,
      \code{=>} and \code{:} tacticals, and intro patterns.}
    \label{fig:inro-rev}
\end{figure}

A typical proof in \lssr is done by moving variables and assumptions
back and forth between the local context and the goal context, placing
the terms to be used in a current proof step on top of the goal
stack.
Such proof context management is done via two complementary
\emph{tacticals}: \code{=>} and \code{:}.
The former one (\code{=>}) moves facts above the sequent line, whereas
the latter one moves facts below, ``pushing'' them to the goal stack.
\autoref{fig:inro-rev1} and~\ref{fig:inro-rev2} showcase the usage of
\code{=>} and \code{:}.
The proof in \autoref{fig:inro-rev1} starts with \code{move=> hA ?},
where \code{move} essentially does nothing: its goal is to serve as a no-op
tactic, to be followed by either \code{=>} or~\code{:}. 
What follows (\code{ => hA ?}) is an example of an \emph{intro
  pattern} (more patterns will be shown below) that introduces the first
two elements of the goal context, the first named \code{hA} and the
second with an auto-generated name.
To apply the introduced assumption \code{hA} we have to revert it back
to the top of the proof context; the next line \code{move: hA} does
exactly that.

Lean users might notice that \code{=>} and \code{:} are similar to the
standard \code{intro} and \code{revert} Lean tactics.
\autoref{fig:inro-rev2} demonstrates the versatility of the \lssr
tacticals when combined with different intro patterns.
%
%
For instance, to swap the first two elements of the proof stack,
\autoref{fig:inro-rev2} makes use of the \code{/[swap]} intro pattern.
This will turn our goal to \code{(@a -> @b) -> @a -> @b}. 
Other useful \ssr-inspired intro patterns include \code{*} for
introducing \emph{all} elements on the proof stack, \texttt{_} for
discarding the first element on the proof stack, and \code{/[dup]} for
duplicating the first element on the stack.
Furthermore, as the purpose of \lssr tacticals is to ``massage'' the
shape of the goal, they can be smoothly combined with other \lssr
tactics that operate directly on the goal stack.
In particular, \autoref{fig:inro-rev2} combines \code{sapply}
with~\code{:} (\ie, \code{sapply: hAiB}), with the effect of first
pushing \code{hAiB : @a -> @b} to the goal stack and then executing
the \code{sapply} tactic, discharging the goal similarly to the example
from \autoref{fig:sapply}.

\subsection{Induction, Case Analysis, and Last-Mile Automation}

\begin{wrapfigure}[7]{r}{0.46\textwidth}
    \vspace{-8pt}
    \captionsetup{aboveskip=5pt}
    \begin{lstlisting}
example (m n : Nat): n <= m -> 
    m - n + n = m := by
    elim: n m=> [| n IHn [| m']] //==
    move=> ?; srw -[2](IHn m') //\end{lstlisting}
    \caption{A proof by mathematical induction}
    \label{fig:elim}
\end{wrapfigure}

\noindent
\lssr combines the tacticals for the goal context management and intro
patterns with tactics for proofs by induction and case analysis.
In particular, \lssr introduces a variant of a standard
\code{induction} tactic called \code{elim}, which expects no explicit
arguments and simply applies the induction principle of
the top element of the goal stack.
As an example, consider the proof by induction in \autoref{fig:elim}
of a simple property of natural addition and subtraction.
As its first step, the proof generalises the two natural constants,
\code{n} and~\code{m} by pushing them onto the goal, making it
\code{@all n m, n <= m -> m - n + n = m};
it is followed by an invocation of the induction principle for natural
numbers on \code{n}, with \code{m} being universally quantified; all
that expressed simply as \code{elim: n m}.

Unlike Lean's \code{induction} tactic, \code{elim} does not introduce
new variables and facts to the local context, leaving those
universally quantified in the respective subgoals.
In our example, using \code{elim} on a natural number introduces two
subgoals, one for the base \code{zero} and one for the inductive step
\code{succ}, with the latter featuring a predecessor variable and an
inductive hypothesis.
Naming variables in subgoals can be done using the \emph{alternation
  separator} intro pattern of the form \text{[ ... | ... | ... ]}
where each \texttt{...} section corresponds to introductions in a
respective subgoal.
We can, thus, revise our proof to be \code{elim: n m=>[| n IHn]},
introducing the variable and the hypothesis step to the local context
as \code{n} and \code{IHn}, respectively.

Let us focus on the second subgoal (\ie, the inductive step), which
now looks as follows:
\vspace{-3pt}
\begin{equation}
\forall m,\, n + 1 \leq m \to m - (n + 1) + (n + 1) = m
\label{eq:goal}
\end{equation}
%
%
The proof of \eqref{eq:goal} can be finished by the case analysis
on~\code{m} using \lssr's \code{scase} tactic, which will generate two
subgoals, one for \code{zero} and one for~\code{succ}. The latter will
introduce yet another universally-quantified variable, which can be
suitably named and moved to the local context: \code{scase=> [| m']}.
Such nested deconstructions of a goal's top variable into cases are so
frequent that \lssr overloads the alternation separator intro pattern
to perform case analyses with simultaneous naming. Hence, we can now
revise our proof to first perform top-level induction on~\code{n} and
then do case analysis on~\code{m} in the resulting second subgoal via
\code{elim: n m=> [| n IHn [| m']]}.

Having executed this proof script line, we are left with three
subgoals to discharge:
%
\begin{align}
  \forall m : \mathbb{N},\, 0 \leq m &\to m - 0 + 0 = m \label{eq:goal2}
  \\
  n + 1 \leq 0 &\to 0 - (n + 1) + (n + 1) = 0 \label{eq:goal3}
  \\
  (n + 1) \leq (m + 1) &\to (m + 1) - (n + 1) + (n + 1) = (m + 1) \label{eq:goal4}
\end{align}
This would be a good time to make use of Lean's famous automation to
solve~\eqref{eq:goal2} and \eqref{eq:goal3}, and simplify
\eqref{eq:goal4}.
\lssr provides the following intro patterns for such ``last-mile''
automation:
\begin{itemize}
\item \code{/=} for simplification by evaluation (Lean's \code{dsimp})
  and \code{/==} for proofs by rewriting (\code{simp})
\item \code{//} for a lightweight automation combining Lean tactics
  \code{trivial} and \code{simp_all}
\item \code{//=} abbreviating \code{dsimp=> //}, and \code{//==}
  abbreviating \code{simp=> //}
\end{itemize}
In this case, using \code{//==} on all generated subgoals leaves us
with only one obligation to prove:
\vspace{-10pt}
\begin{equation}\label{eq:goal5}
n \leq m' \to m' - n + (n + 1) = (\gbm{m'} + 1)
\end{equation}

\vspace{-10pt}

\subsection{Targeted Rewriting using Equality Hypotheses}

\noindent
The remaining goal~\eqref{eq:goal5} can be proven using the previously
introduced induction hypothesis \code{IHn : @all (m : Nat), n ≤ m → m
  - n + n = m}. 
Intuitively, the proof can be completed by replacing the
\graybox{highlighted} \emph{second} occurrence of $m'$ in
\eqref{eq:goal5} with $m' - n + n$ using the right-hand side of the
equality in \code{IHn}, and then finish the proof by the properties of
linear arithmetic and reflexivity.
This can be achieved by \lssr rewriting tactic \code{srw} allowing one
to specify the term occurrence to be rewritten.
The proof script in \autoref{fig:elim} ends with moving the inequality
\code{n ≤ m'} to the local context via \code{move=>?}, followed by
\code{srw -[2](IHn m') //}, where \code{-} stands for the
right-to-left rewrite direction and \code{[2]} denotes the specific
occurrence of \code{m'}.
Performing this rewrite generates the obligation \code{n ≤ m'}, which
is discharged via \code{//}.

The \code{srw} tactic generalises Lean's vanilla \code{rw}, allowing
for constructions such as:
\vspace{-3pt}
\[
\text{\code{srw (drop_nth _ lt_i_m) //== -[1]h nth_index //
    -index_mem}}
\]
\vspace{-3pt}
that interleave rewrites with simplifications, resolving the appearing
subgoals with~\code{//}.

\vspace{-5pt}

\subsection{Fine-Grained Proof State Exploration and Error
  Highlighting}
\label{sec:debug}

\noindent
As our examples so far demonstrate, proofs in \lssr tend to be quite
compact, thanks to the concise nature of the tactic language that
chains multiple manipulations with the goal and the proof context into
a single line of a proof script.
The downside of this style is that proofs often become difficult to
follow and refactor---a frequent complaint by newcomers to
Coq/\ssr who are familiar with the vanilla Coq proof mode.
The problem of proof script understanding and maintenance in \ssr is
exacerbated by the fact that Coq can only display the proof context
between \emph{complete} series of chained tactic applications
separated by periods.
When a mistake is made in the middle of a line in an \ssr
proof script, one typically has to manually break the line into
period-terminated parts, performing a binary search-style debugging.
In addition to that, if a particular intro pattern cannot be applied
in \ssr, say we write \code{[| | m]} with three alternations instead
of two in the proof of \autoref{fig:elim}, Coq will report an error
for the entire script line, further complicating the search for a fix.


\begin{figure}[t]
    \begin{subfigure}{0.49\textwidth}
        \fbox{\includegraphics[width=0.95\textwidth]{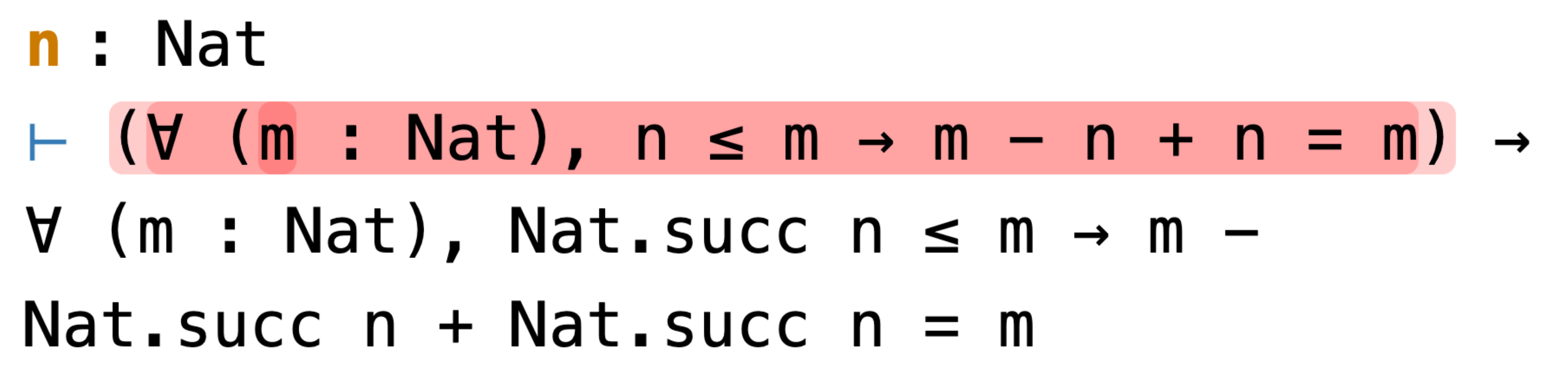}}
        \caption{The proof view before introducig \code{IHn}}
        \label{fig:before-IHn}
    \end{subfigure}
    \begin{subfigure}{0.49\textwidth}
        \fbox{\includegraphics[width=0.945\textwidth]{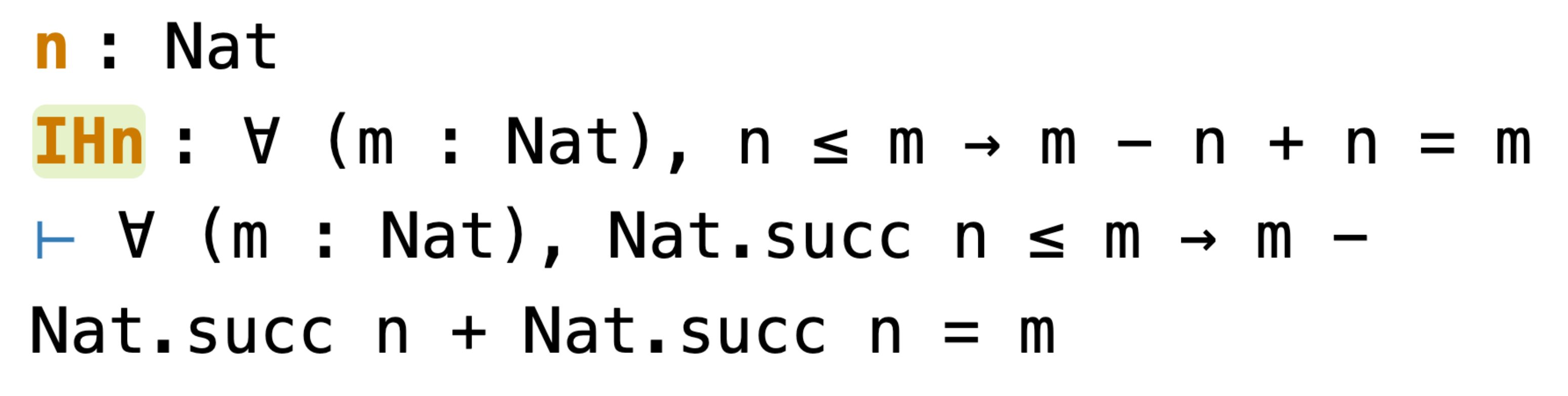}}
        \caption{The proof view after introducig \code{IHn}}
        \label{fig:after-IHn}
    \end{subfigure}
    \captionsetup{aboveskip=5pt}
    \captionsetup{belowskip=-10pt}
    \caption{Visual support for fine-grained proof state exploration
      with \lssr proof scripts.}
\end{figure}

\lssr overcomes these hurdles, providing an improved (compared to
\ssr) experience to the users, thereby substantiating the usability
claim~(\ref{label:ui}) from \autoref{sec:intro}.

First, the better proof debugging is made possible by the powerful
feature of Lean that allows the DSL designer to annotate each symbol
in the tactic definition with the proof state before and after its
execution.
For instance, if the user puts the pointer in the text buffer with her
proof script before the \code{IHn} token in \autoref{fig:elim}, the
goal display will be as in \autoref{fig:before-IHn}, highlighting the
upcoming change in the goal in red.
For the pointer positioned right after \code{IHn}, the goal display
shows the proof context as in~\autoref{fig:after-IHn}.
Second, Lean provides a tunable mechanism for precise error
highlighting, so if we write \code{[| | m]} with three alternations
instead of two, only highlight the \code{[| | m]} part of the script
will be marked as an error.
%


\vspace{-5pt}

\subsection{Forward Reasoning via Views}
\label{sec:views}

\begin{wrapfigure}[8]{r}{0.46\textwidth}
\vspace{-8pt}
\captionsetup{aboveskip=2pt}
\begin{lstlisting}
example (A B C : Prop) :
  (A → B) → (B → C) → A → C :=
  by intro AiB BiC Ha
     apply BiC
     apply AiB
     exact Ha\end{lstlisting}
\caption{A backward proof in vanilla Lean}
\label{fig:back}
\end{wrapfigure}
\noindent
So-called \emph{backward} proofs, in which a conclusion of a goal is
gradually strengthened to be eventually discharged from the available
hypotheses or axioms, are prevalent in interactive proofs.
A very simple example of such a proof is given in \autoref{fig:back},
which proves the conclusion~\code{C} using the reverse-order
applications of the assumptions \code{BiC}, \code{AiB}, and,
eventually, \texttt{Ha~:~A}.

\begin{wrapfigure}[8]{r}{0.46\textwidth}
\vspace{-8pt}
\captionsetup{aboveskip=2pt}
\begin{lstlisting}
example (A B C : Prop) :
  (A → B) → (B → C) → (A → C) :=
  fun AiB BiC Ha ↦ 
    have Hb : B := AiB Ha
    have Hc : C := BiC Hb
    show C from Hc\end{lstlisting}
\caption{A forward proof in vanilla Lean}
\label{fig:forw}
\end{wrapfigure}
That said, proof engineers trained in a tradition of purely
mathematical proofs often prefer the complementary \emph{forward}
style, in which the assumptions are gradually weakened to eventually
imply the conclusion~\cite[Chapter~4]{love}.
A typical forward proof in Lean follows the so-called
\emph{assume}/\emph{have}/\emph{show} pattern, which explicitly states
the $\forall$-quantified variables and hypotheses as parameters of the
proof term (\emph{assume}) and the proof is comprises a number of
established auxiliary facts (\emph{have}) that make the derivation of
the conclusion trivial (\emph{show}).
An example of such a forward proof in the conventional Lean notation
is shown in \autoref{fig:forw}.

\lssr offers an alternative way to construct forward proofs using the
idea of \emph{view intro patterns} or simply \emph{views}, adopted
from the original \ssr~\cite{Gonthier-al:TR}.
The proof of the fact from \autoref{fig:back} and~\ref{fig:forw} can
be achieved in \lssr by the following single line:
\vspace{-3pt}
\begin{center}
\code{by sby move=> AiB BiC /AiB/BiC}  
\end{center}
\vspace{-3pt}
Ignoring \code{sby} for a moment, notice that the proof first moves
the hypotheses \code{AiB : A → B} and \code{BiC : B → C} to the
context.
What follows is an application of \code{AiB} to the leading (\ie, top)
assumption of the goal \code{A}, which ``switches'' it to \code{B},
making the goal to be \code{B → C}.
The subsequent view \code{/BiC} switches the leading assumption from
\code{B} to \code{C}.
The remaining goal \code{C → C} can be discharged by \code{trivial}
(or \lssr's \code{//}).
This is done by the \code{sby} tactical at the beginning of the line,
which either completes the proof via \code{//} or fails.
In fact, this makes the second view redundant, so the proof can be
shortened to \code{by sby move=> AiB ? /AiB}.

\vspace{-5pt}

\subsection{Proofs by Computational Reflection}
\label{sec:refl}

The \lssr demonstration in \autoref{sec:move}--\ref{sec:views} was
merely a build-up to the true essence of the framework: mechanised
reasoning that features both \emph{deduction} and \emph{proof via
  computation} by manipulating with equivalent \emph{logical} and
\emph{symbolic} representations of decidable facts.

\begin{figure}[t]
\begin{tabular}{c}
\begin{subfigure}{0.5\textwidth}
\captionsetup{aboveskip=0pt}
\begin{lstlisting}
inductive even : Nat → Prop 
  where
  | ev0 : even 0
  | ev2 : ∀ n, even n → even (n + 2)
\end{lstlisting}
\caption{Even numbers as a logical predicate}
\label{fig:even}
\end{subfigure}
\begin{subfigure}{0.49\textwidth}
\captionsetup{aboveskip=0pt}
\begin{lstlisting}
def evenb : Nat → Bool
  | 0     => true 
  | 1     => false
  | n + 2 => evenb n
\end{lstlisting}
\caption{Even numbers via a Boolean function}
\label{fig:evenb}
\end{subfigure}
\\[5pt]
\begin{subfigure}{\textwidth}
\captionsetup{aboveskip=0pt}
\begin{lstlisting}
@[reflect] instance evenP n : Reflect (even n) (evenb n) := ...
#reflect even evenb

example n m : even n → even (m + n) = even m := by 
  elim=> // n' _ <-
  srw -Nat.add_assoc /==
\end{lstlisting}
\caption{A proof using reflection between the logical (\fcode{even})
  and symbolic (\fcode{evenb}) definition of even numbers.}
\label{fig:reflect}
\end{subfigure}
\end{tabular}
\captionsetup{aboveskip=5pt}
\captionsetup{belowskip=-10pt}
\caption{Two equivalent definitions of even numbers (top), and a proof
  by reflection (bottom).}
\label{fig:refl}
\end{figure}

As a motivating example showing how to unify both these proof styles
in \lssr, consider the following simple proposition: \emph{assuming
  $n$ is even, $n + m$ is even if and only if $m$ is even}.
How should we express a decidable property, such as evenness?
Proof assistants based on higher-order logic, such as Coq,
Isabelle/HOL, and Lean provide two conceptual ways to do so: as a
\emph{predicate} (\cf~\autoref{fig:even}) and as a \emph{function}
(\cf~\autoref{fig:evenb}).
The former style comes with an advantage of providing a richer
knowledge when a property occurs in a goal's \emph{assumption}, \eg,
by assuming \code{even n} we can deduce that it's either zero or
another even number plus two.
The latter comes with a set of reduction/rewrite principles, such as, \code{evenb n + 2 = evenb n}, which is advantageous for
simplifying the \emph{conclusion} of a goal.

One can indeed establish an equivalence between the definitions
\code{even} and \code{evenb}, proving that \code{@all n, even n ↔
  (evenb n = true)}.
Such an equivalence, when expressed as an instance of the
\code{Decidable} type class in Lean, can be exploited by allowing the
user to define computations that explicitly feature a predicate
instead of its computable version, such as:
\begin{center}
\code{def even_indicator (n: Nat) : Nat := if even n then 1 else 0}  
\end{center}
This treatment of decidability also explains the somewhat frivolous
phrasing of our proposition of interest in terms equality instead of
bi-implication (\code{<->}) in the statement in \autoref{fig:reflect}.
To summarise, for \code{Decidable} properties, Lean aggressively
promotes the logical definitions to be considered as ``primary'',
while the symbolic ones serving merely as helpers for defining
computations and, occasionally, for simplifying goals in the proofs
about equivalent predicates.

This is exactly where \lssr comes into play, making such simplifications
transparent to the user. Consider the proof of the example in
\autoref{fig:reflect}.
The first line initiates the induction on the \code{even n} premise,
discharging the first trivial subgoal (for the case \code{ev0}),
binding the first argument \code{n'} of the case \code{ev2}, ignoring
the \code{even n'} assumption (via \texttt{_}), and performing the
right-to-left rewrite using the induction hypothesis \code{even (m +
  n') = even m} in the conclusion (via the \code{<-} intro pattern),
leaving us with the following goal to prove:
\vspace{-3pt}
\begin{center}
\code{even (m + (n' + 2)) = even (m + n')}  
\end{center}
\vspace{-3pt}
The remaining goal is not difficult to prove by using associativity of
addition and showing that \code{@all x, even (x + 2) = even x}.
But also, this is an equality that we should have ``for free'', as it
can be derived from the last clause of the definition of \code{evenb}
in \autoref{fig:evenb}!

To account for scenarios like this one, \lssr provides a machinery
that allows one to inherit reduction rules of a recursive (symbolic)
boolean definition \code{b} to simplify instances of the equivalent
logical proposition \code{P}.
To achieve that, one must prove an instance of the \code{Reflect P
  b} predicate, which is essentially equivalent to proving \code{P ↔
  (b = true)} (we postpone the details of the \code{Reflect}
definition until \autoref{sec:implementation}) and register this
equivalence using the pragma \code{#reflect P b}, which is exactly
what is done by the first two lines of \autoref{fig:reflect}.
In this case, doing so will add three simplification rules for
\code{even} into a database of rewrites used by simplification tactics
such as \code{simp} and the corresponding \lssr patterns, \eg,
\code{/==}:
\begin{center}
\code{even 0 = True} \quad\quad  
\code{even 1 = False} \quad\quad  
\code{@all n, even (n + 2) = even n}
\end{center}
%
%
These equalities allow us to finish the proof by \code{/==}, rewriting
\code{even} \emph{as if it were} \code{evenb}.

\lssr's \code{Reflect} type class is well integrated with the current Lean
ecosystem. Once one has proven \code{Reflect P b}, the corresponding
\code{Decidable} instance is automatically derived for~\code{P}.
%
%
Readers familiar with \ssr for Coq could notice that a proof of the
statement from \autoref{fig:reflect} in it would be more verbose, as
it would require an \emph{explicit} switching between the logical
predicate and its symbolic counterpart in the conclusion by
\texttt{apply}ing a special \code{Reflect}-lemma to the goal.
This is not necessary in \lssr thanks to Lean's ability to use the
derived rewriting principles (\eg, the three above for \code{even})
directly on logical representations.
We believe this example supports the expressivity claim
(\ref{label:express}) from \autoref{sec:intro}.

\vspace{-5pt}

\subsection{Putting It All Together}
\label{sec:putting-it-all}

We conclude this tutorial by showing how all the features of \lssr
introduced in \autoref{sec:move}--\ref{sec:refl} work in tandem in a
proof of an interesting property: transitivity of the subsequence
relation on lists of elements with decidable equality.
\autoref{fig:subseq} presents two ways to define the relation.
The first one is via a predicate and an auxiliary \code{mask}
function. The \code{mask} function takes two lists: a list of Boolean
values~\code{m} and a list \code{s} of elements of some type
\code{@a}. 
For each element of \code{s}, \code{mask} removes it if an element in
\code{m} at the same position is either \code{false} or not
present (\ie, \code{m} is shorter than \code{s}).
Then the representation of subsequence as a logical predicate states
that~\code{s1} is a subsequence of~\code{s2}, if \code{mask m s2 = s1}
for some mask~\code{m}.
The second definition is via a total recursive function \code{subseqb}
returning a Boolean. 
This definition we checks that each element in the first list
corresponds to some element in the second list using the decidable
equality on the values of type~$\alpha$, whose existence is postulated
in the first line of the listing.
For the propositional and Boolean representations above, we can prove
a \code{Reflect} instance (\cf~\autoref{sec:refl}), and transport
reduction principles of \code{subseqb} to \code{subseq} by adding the
\code{#reflect} pragma as at the line~1 of \autoref{fig:trans}.

\begin{figure}[t]
\begin{subfigure}{0.55\textwidth}
\begin{lstlisting}[basicstyle=\footnotesize\ttfamily]
variable {α : Type} [DecidableEq α]

def mask: List Bool → List α → List α
  | b :: m, x :: s => 
    if b then x :: mask m s else mask m s
  | _, _ => []

def subseq (s1 s2 : List α) : Prop := 
  ∃ m, length m = length s2 /\ 
    s1 = mask m s2

def subseqb: List α → List α → Bool
  | [], _ :: _          => true
  | s, []               => s = []
  | s@(x :: s'), y :: r => 
      subseqb (if x = y then s' else s) r

\end{lstlisting}
\caption{Two representations of the subsequence relation}
\label{fig:subseq}
\end{subfigure}
\hspace{-3mm}
\begin{subfigure}{0.45\textwidth}
\begin{lstlisting}[basicstyle=\footnotesize\ttfamily,numbers=left,numberstyle=\tiny\color{ACMRed}]
#reflect subseq subseqb

def transitive (R: α → α → Prop) :=
  ∀x y z, R x y → R y z → R x z

example: transitive (@subseq α) := by
  move=> ?? s ![m2 _ ->] ![m1 _ ->]
  elim: s m1 m2=> [// |x s IHs1]
  scase=> [// | [] m1 /= m2]
  { -- m1's head is false
    scase!: (IHs1 m1 m2)=> m sz_m ->
    sby exists (false :: m) }
  -- m1's head is true  
  scase: m2=> [|[] m2] //=
  scase!: (IHs1 m1 m2)=> m sz_m ->
  sby exists (false :: m)
\end{lstlisting}
\caption{Proof that \code{subseq} is transitive}
\label{fig:trans}
\end{subfigure}
\captionsetup{belowskip=-10pt}
\caption{A subsequence relation and the proof of its transitivity in
  \lssr.}
\label{fig:subrefl}
\end{figure}

\autoref{fig:trans} shows the entire transitivity proof.
At line~7, it introduces a new intro pattern \code{![...]}, which is
an advanced version of the case analysis \code{[...]}: not only does
it destruct the structure on top of the proof stack, but also does so
for its nested structures.
For instance, the second occurrence of \code{![...]} will turn the
goal \code{(∃m, length m = length s ∧ w = mask m s) → ...} into
\code{@all m, length m = length s → w = mask m s → ...} by destructing
both the existential quantifier (\ie, the dependent product) and its
nested conjunction.
The remaining in-place rewrites (via \code{->}) at the line~7 turn the
goal into:
\begin{equation}
\text{\code{subseq (mask m2 (mask m1 s)) s}}
\label{eq:goal-trans}
\end{equation}
In plain text, we have to show that if we apply \code{mask} to some
sequence~\code{s} with two arbitrary masks,~\code{m1} and~\code{m2},
the resulting sequences would be a subsequence of~\code{s}.
Line~8 advances the proof by induction on~\code{s}, after generalising
the goal over \code{m1} and \code{m2} and discharging the base case
with \code{//}, which implicitly uses the rewrites allowed by the
definition of \code{mask}.%

Next, line~9 performs the case analysis on \code{m1}. When it is empty, goal
\eqref{eq:goal-trans} becomes \code{subseq [] s}, which can be automatically
reduced to \code{True} by employing the reflection between \code{subseq} and
\code{subseqb}.
The case when \code{m2} is empty is handled automatically in a similar
fashion at line~14.
We are now left with the following two remaining goals:

\begin{itemize}[topsep=1pt]
\item If \code{m1} is non-empty and its first element is \code{false} (lines
  11-12), then after all simplifications, goal~\eqref{eq:goal-trans}
  is reduced to \code{subseq (mask m2 (mask m1 s)) (x :: s)},
  where \code{x} is a head of the initial list \code{s}. 
  Here we employ the induction hypothesis \code{IHs1} at line~11, to
  state that \code{mask m2 (mask m1 s)} is a subsequence of \code{s},
  which means that \code{mask m2 (mask m1 s)} is just \code{mask m s},
  for some mask \code{m}.
  To finish the proof we instantiate the existential quantifier in the
  definition of \code{subseq} by taking \code{false :: m}, which, when
  applied to \code{x :: s} produces exactly \code{mask m s} (line~12).
\item If the first element of \code{m1} is \code{true}, we perform case analysis
  on \code{m2} (line~14), dispatching the empty case in a way similar to dealing
  with \code{m1} at line~9.
  When \code{m2}'s first element is also \code{true} is
  trivial as in this case neither of the masks removes the head of
  \code{s} and our goal will simplify to \code{subseq (x :: mask
    m2 (mask m1 s)) (x :: s)} and then, by reflection, to \code{subseq
    (mask m2 (mask m1 s)) s}, which is trivially implied by the
  induction hypothesis.
  All this is done by \code{//=} at the end of line~14.
  Finally, the case when \code{m2}'s head is \code{false}
  (lines~15-16) is similar to the proof at lines~11-12.
\end{itemize}

\noindent
The proof in \autoref{fig:subrefl} comes from our case study:
migrating the MathComp library \code{seq} from Coq/\ssr to Lean.
In \autoref{sec:ssrseq} we will further elaborate on this
mechanisation effort.

\vspace{-5pt}

\section{Implementing \lssr using Lean Metaprogramming}
\label{sec:implementation}

In this section we shed light on the implementation of \lssr, which
makes extensive use of Lean's metaprogramming facilities.
Despite the abundance of available tutorials on implementing tactics
in Lean, it still took significant effort to understand how to compose
different features to achieve the desired behaviour,
following examples from the Lean Metaprogramming Book~\cite{metalean},
several research papers \cite{selsam2020tabled,Ullrich2022}, Lean
source code, and online meetings with Lean developers.
This makes us believe that our report on this experience might be of
interest even for seasoned Lean users who consider developing their
own tactics.

Besides the new \lssr-specific tactics, such as \code{elim},
\code{scase}, \code{sapply}, \etc., the three essential enhancements
made by \lssr over the vanilla Lean tactic language are:
\begin{enumerate}[topsep=2pt]
    \item \emph{Intro patterns}: the set of commands that can follow \code{=>}
    \item \emph{Rewrite patterns}: the set of commands that can follow \code{srw}
    \item \emph{Revert patterns}: the set of commands that can follow \code{:}
\end{enumerate}
We refer to those commands as patterns because they are meant to
(partially) \emph{match} some part of the goal. 
The first set, \ie, intro patterns, includes \code{?}, \code{*}, \code{[..
  | ... | ..]}, \etc. 
The second set, rewrite patterns, includes \code{eq}, \code{[i j
  ...]eq}, \code{-[i j ...]eq}, \etc, where \code{eq} is an arbitrary
equality fact. 
For the sake of space, we do not detail revert patterns in this paper,
but they are fully documented in the \texttt{README} file of the
accompanying code repository~\cite{lssr}.

The characterisation of the three pattern categories is non-exclusive:
some \lssr commands, \eg, \code{//} and \code{/==}, can be used
both as intro patterns (following \code{=>}) and rewrite patterns
(following \code{srw}).
To avoid code duplication, those ``last-mile automation'' commands
themself form a set of \emph{automation patterns} that are implemented
separately and then registered to work in positions of both intro and
rewrite patterns.

\begin{wrapfigure}[7]{r}{0.58\textwidth}
\includegraphics[width=0.58\textwidth]{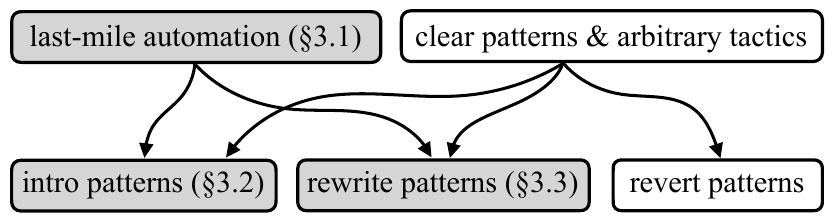}
\captionsetup{aboveskip=8pt}
\caption{High-level structure of \lssr}
\label{fig:lssr}
    \end{wrapfigure}
An informal depiction of \lssr patterns can be found in
\autoref{fig:lssr}. Each box stands for a separate set of pattens, the
arrows denote the dependencies between those sets. 
In \autoref{sec:auto}--\ref{sec:srw}, we discuss the implementation of
the components highlighted by grey boxes, concluding with a discussion
on \lssr support for computational reflection in
\autoref{sec:ssrimpl}.
We do not assume the reader to be familiar with metaprogramming in
Lean and will explain the required notions as we progress with our
presentation.

\vspace{-10pt}

\subsection{Syntax, Macros, and Elaboration for the Last-Mile Automation}
\label{sec:auto}

We start by discussing the three key features of Lean metaprogramming
essential for our \lssr embedding: (1)~syntactic categories,
(2)~macro-expansion, and (3)~elaboration rules, using the
implementation of automation patterns as an example.
The implementation is shown in \autoref{fig:ssrTriv}. At the high
level, we first define a syntax for a sequence of such patterns (lines
1-8), then tell Lean how to execute such a sequence (lines 10-22),
concluding with the definition of \code{sby} tactical.
More detailed explanation of the code follows below.

\begin{figure}[h]
    \centering
\begin{subfigure}{0.45\textwidth}
\begin{lstlisting}[basicstyle=\scriptsize\ttfamily,numbers=left,numberstyle=\tiny\color{ACMRed}]
declare_syntax_cat ssrTriv
syntax "//" : ssrTriv
syntax "/=" : ssrTriv
syntax "/==" : ssrTriv
syntax "//=" : ssrTriv
syntax "//==" : ssrTriv
declare_syntax_cat ssrTrivs
syntax (ppSpace colGt ssrTriv)* : ssrTrivs

elab_rules : tactic
    | `(ssrTriv|  /=) => 
    run `(tactic| try dsimp)
    | `(ssrTriv| /==) => 
    run `(tactic| try simp)
    | `(ssrTriv|  //) => ... -- omitted 
\end{lstlisting}
\end{subfigure}
\hspace{4mm}
\begin{subfigure}{0.45\textwidth}
    \begin{lstlisting}[basicstyle=\scriptsize\ttfamily,firstnumber=16,numbers=left,numberstyle=\tiny\color{ACMRed}]
macro_rules
    | `(ssrTriv| //= ) => `(ssrTrivs| /= // )
    | `(ssrTriv| //==) => `(ssrTrivs| /== //)

elab_rules : tactic
    | `(ssrTrivs| $ts : ssrTriv *) => 
        for t in ts do allGoals $ evalTactic t

elab sby:"sby " ts : tacticSeq : tactic => do
  evalTactic ts
  unless (<- getUnsolvedGoals).length = 0 do
    allGoals $ run `(ssrTriv| //)
  unless (<- getUnsolvedGoals).length = 0 do
    throwErrorAt sby "No applicable tactic"
    \end{lstlisting}
    \end{subfigure}
    \captionsetup{aboveskip=2pt}
    \captionsetup{belowskip=-10pt}
    \caption{Implementing of last-mile automation in \lssr using Lean
      metaprogramming.}
    \label{fig:ssrTriv}
\end{figure}

Line 1 defines a \emph{syntactic category} called \code{ssrTriv} for
the \lssr automation patterns. One can think of a syntactic category as
of a set of syntax expressions grouped together under a single name, for
which interpretations can be given.
Next, lines 2-6 define elements of this category: \code{//},
\code{/=}, \code{/==}, \code{//=} and \code{//==}.
At line~7, we define yet another syntax category \code{ssrTrivs} for
sequencing automation patterns.
Elements of this category are iterated sequences of \code{ssrTriv}
separated with a space~\code{ppSpace}.
To ensure that Lean does not try to parse the next line after a
sequence of the patterns \code{ssrTriv} as a its continuation, we also
insert the \code{colGt} element, which allows for line breaks, but
only if the column number of the pattern on a new line is greater than
the column number of the last pattern on the line above, in the spirit
of Landin’s Offside Rule~\cite{Landin66}.

Having defined the syntax for the automation pattern, we proceed
to ascribe semantics to it. 
Lines 10-15 give a meaning to basic automation patterns by using
Lean's standard mechanism of elaboration rules via the
\code{elab_rules} directive. 
Each elaboration rule maps a piece of a syntax into an execution
inside Lean's \code{TacticM} monad~\cite[\S{8.3}]{love},
%
%
thus defining a tactic in terms of other Lean tactics, with
access to the global proof state.
For example, \code{/=} is elaborated into executing the standard
\code{try dsimp} tactic (lines~11-12).
The remaining automation patterns, \code{//=} and \code{//==} are
defined in terms of the more primitive elements of \code{ssrTriv}.
This is done using Lean's rules for macro-expansions, following the
\code{macro_rules} directive.
Macro-expansion rules in Lean map elements of syntax into executions
within the~\code{MacroM} monad, returning another element of syntax as a
result.
Those rules are applied by the Lean interpreter after the parsing
stage, but before the elaboration stage.
Here, we simply expand \code{//=} to \code{/= //}, and \code{//==} to
\code{/== //}, as shown at lines~16-18 of \autoref{fig:ssrTriv}.


Since both \code{//=} and \code{//==} are represented as elements of
the \code{ssrTrivs} category, we have to tell Lean how to elaborate
them as well.
For the sake of demonstration, we are going to do it by employing
one of the most powerful tools in Lean's metaprogramming toolbox: the
\code{evalTactic} directive.
This directive takes an arbitrary piece of syntax and tries to
elaborate it into a sequence of tactics to be executed, using a set of
macro-expansion and elaboration rules available in the \emph{dynamic}
context, \ie, the rules available at the \emph{usage site} in a
particular proof script.
%
%
%
We will get to experience \code{evalTactic} in its full glory in
\autoref{sec:intropats}. For now, let us highlight one of its most
useful features: annotating each token in the given syntax with a
correspondent proof state.
That is, when elaborating and executing its argument,
\code{evalTactic} saves the goal before and after executing each
token, 
%
allowing for the interactive fine-grained proof state exploration we
presented in~\autoref{sec:debug}.
%

Finally, lines~24-30 implement the \code{sby} tactical
(\autoref{sec:views}). Line~24 defines both its syntax and elaboration
using the \code{elab} directive.
The tactical starts with the string ``sby'' (its position in a
concrete syntax tree is bound as \code{sby}), followed by a sequence
of tactics \code{ts}.
We first run this tactic sequence (line~15), then, unless it has
solved all goals, we run \code{//} (lines~26-27).
Otherwise, if there are unsolved goals, an error is reported at the
\code{sby} position.

\vspace{-5pt}

\subsection{Intro Patterns, Modularity, and Extensibility}
\label{sec:intropats}

Let us now briefly go through the high-level structure of
the implementation of intro patterns. 
While to a large extent similar to that of last-mile automation,
this part of our development showcases two new noteworthy aspects of
\lssr internals.
First, we will demonstrate the \emph{modular} nature of pattern
definitions by showing how the intro pattens incorporate those for
automation.
Second, we will show how the intro patterns that come with \lssr can
be easily \emph{extended} for domain-specific proofs, thus addressing
claim~(\ref{label:ext}) from \autoref{sec:intro}---an aspect of our
implementation made possible by the dynamic nature of the
\code{evalTactic} directive.

\begin{figure}[h]
    \centering
\begin{subfigure}{0.45\textwidth}
\begin{lstlisting}[basicstyle=\scriptsize\ttfamily,numbers=left,numberstyle=\tiny\color{ACMRed}]
declare_syntax_cat ssrIntro
syntax ssrTriv : ssrIntro
syntax ident : ssrIntro
syntax "?" : ssrIntro
...
declare_syntax_cat ssrIntros
syntax (ppSpace colGt ssrIntro)* : ssrIntros

elab_rules : tactic
  | `(ssrIntro| $t:ssrTriv) => evalTactic t
  | `(ssrIntro|$i:ident) => 
    run `(tactic| intro $i:ident)
  | `(ssrIntro| ?) => 
    run `(tactic| intro _)
  ...
\end{lstlisting}
\end{subfigure}
\hspace{4mm}
\begin{subfigure}{0.45\textwidth}
    \begin{lstlisting}[basicstyle=\scriptsize\ttfamily,firstnumber=16,numbers=left,numberstyle=\tiny\color{ACMRed}]
elab_rules : tactic
  | `(ssrIntros| $is:ssrIntro*) => 
    for i in is do allGoal $ evalTactic i

elab t:tactic "=> " is:ssrIntros : tactic => 
  do evalTactic t; evalTactic is
...

macro_rules 
  | `(ssrTriv| //) => `(tactic| omega)
...
syntax "!" : ssrIntro
elab_rules : tactic
  | `(ssrIntro| !$i:ident) => 
    run `(tactic| ext)\end{lstlisting}
\end{subfigure}
\captionsetup{aboveskip=2pt}
\captionsetup{belowskip=-10pt}
\caption{Implementation of \lssr intro patterns and their extensions.}
\label{fig:ssrIntro}
\end{figure}

\noindent
\autoref{fig:ssrIntro} presents a fragment of the implementation of
intro patterns for \lssr.
Following the template outlined in \autoref{sec:auto}, it defines the
syntax for a single intro pattern (lines 1-6) and multiple intro patterns
(lines 5-7). 
%
%
Now, to allow for last-mile automation within intro patterns, we only
have to explicitly add line~10, which, intuitively, says: once
you meet a \code{ssrTriv} inside some \code{ssrIntro}, just handle it
with the already defined macro/elaboration rules.
Lines~11-14 provide examples of an intro pattern elaboration rule: a
constant introduction with an explicit name~\code{i} (lines~11-12) and
with an autogenerated name (lines~13-14), handled using Lean's
\code{intro} tactic.
Lines~16-18 handle a sequence of intro patterns.
Finally, a new tactic \code{=>} is defined at lines~20-21 to allow use of
intro pattens in a proof script.

Lines 24-25 of \autoref{fig:ssrIntro} show the first example of
extensibility afforded by Lean metaprogramming: augmenting the
automation pattern \code{//} to discharge arithmetic facts by calling
the standard \code{omega} tactic.
This definition \emph{enhances} the behaviour of the pattern in the
entire scope below the macro-expansion rule (keep reading to learn why
it does not \emph{replace} the original behaviour of \code{//}).
Another example of extensibility is presented at lines~27-30,
integrating a domain-specific tactic into \lssr patterns.
The mathlib's \code{ext} tactic automaticallly applies extensionality
axioms, \eg, turning a goal of the form \code{f = g}, where both
\code{f} and \code{g} are functions, into \code{@all x, f x = g x},
followed by introduction of~\code{x}.
We integrate it into \lssr intro patterns by definining the syntax
\code{!} for it as a new \code{ssrIntro} pattern at line~27, and by
elaborating the \code{!} pattern to the \code{ext} tactic at lines
28-30.
%

The desired effect in both scenarios outlined above is achieved by the
two somewhat non-obvious practices we follow in our implementation of
\lssr patterns:
\begin{enumerate}[(a),topsep=2pt]
\item The semantics of the patterns is implemented exclusively using
  the \code{macro_rules} and \code{elab_rules} commands, while their
  evaluation is \emph{always} done using \code{evalTactic}, which is
  invoked from the definitions of pattern sequences, as appear in
  \autoref{fig:ssrTriv} as \autoref{fig:ssrIntro}.
\item Elaboration rules for each particular pattern are defined
  separately from others. 
\end{enumerate}

\noindent
Following these practices makes it possible to extend \lssr with new
\code{macro_rules} and \code{elab_rules} in a modular fashion.
To extend \lssr with a new pattern or add a new behaviour, it suffices
to just add a new elaboration/macro rule for it.
The reason why it works follows from the way \code{evalTactic}
performs the bookkeeping of the macro/elaboration rules and their
application.
Any new such rule is added at the top of a \emph{rule database stack}.
Whenever \code{evalTactic} is invoked (\eg, when elaborating and
executing a particular sequence of intro patterns as per line~18 of
\autoref{fig:ssrIntro}), it will go through the stack of the saved
rules, starting from the top, until it either finds the first one that
executes without errors (\ie,~without throwing an exception as in
\autoref{fig:ssrTriv} at line~29), or exhausts the stack.%
\footnote{It is a good practice to put elaboration into tactics, which
  either fully solve the goal or fail otherwise, closer to the top of
  the stack, so they would not be preempted by elaboration into
  tactics that never fail.}

In particular, in case of \code{//} supplied with the rule at the
lines~24-25 of \autoref{fig:ssrIntro}, \code{evalTactic} will first
try to expand it into \code{omega}.
If \code{omega} executes without errors (this is only possible if it
solves the goal completely), \code{evalTactic} looks no further.
If \code{omega} fails, leaving the goal unchanged, \code{evalTactic}
proceeds with the next option on the rule stack, \ie the one at
line~15 of \autoref{fig:ssrTriv} (assuming no other registered rules
for \code{//}).
%
%
This is the reason why lines~24-25 of
\autoref{fig:ssrIntro} \emph{do not alter} the behavior of \code{//}
but augment it with \code{omega}.

\vspace{-8pt}

\subsection{Rewrite Patterns and Custom Environment Extensions}
\label{sec:srw}

The implementation of \lssr rewrite patterns relies on yet another
very useful feature of Lean metaprogramming: \emph{custom environment
  state extensions}.
%
%
The need for environment state extensions comes from the rather
restrictive format of \code{evalTactic}, which takes only one
argument: the syntax of the tactic to elaborate and run.
Therefore, the execution of a tactic implemented via \code{evalTactic}
can only depend on the general proof environment, \ie, the proof
context and the set of all available definitions available at the
point of its invocation.
So far, this has been sufficient for our needs to implement patterns
for automation and introduction that were ``context-independent''.
The behaviour of a useful tactic for rewriting is a bit more complex,
since this tactic is expected to take a rewrite \emph{location}, \ie,
a local hypothesis or the goal (its default behaviour).
While the location of a rewrite is accessible at the level of the
entire tactic command, it cannot be directly passed to
\code{evalTactic} when elaborating each individual rewrite pattern, as
the pattern itself is \code{evalTactic}'s sole argument.
Custom environment state extensions solve this issue by providing
access to \emph{additional state} that can be used to store such
data when elaborating a top-level command, making its accessible for
the elaboration rules of its inner components (\ie, the patterns).

\begin{figure}[t]
    \centering
\begin{subfigure}{0.42\textwidth}
\begin{lstlisting}[basicstyle=\scriptsize\ttfamily,numbers=left,numberstyle=\tiny\color{ACMRed}]
declare_syntax_cat ssrRw
syntax ssrTriv : ssrRw
syntax term:max : ssrRw
declare_syntax_cat ssrRws
syntax (ppSpace colGt (ssrRw))* : ssrRws

abbrev LocationExtState := Option 
  (TSyntax `Lean.Parser.Tactic.location)

initialize locExt : 
  EnvExtension LocationExtState ←
  registerEnvExtension (pure none)
\end{lstlisting}
\end{subfigure}
\hspace{4mm}
\begin{subfigure}{0.48\textwidth}
    \begin{lstlisting}[basicstyle=\scriptsize\ttfamily,firstnumber=16,numbers=left,numberstyle=\tiny\color{ACMRed}]
elab_rules : tactic
  | `(ssrRw| $t:ssrTriv) => 
    evalTactic t
  | `(ssrRw| $i:term) => do
    let l <- locExt.get
    run `(tactic| rw [$i:term] $(l)?)
...

elab "srw" rs:ssrRws l:(location)? : tactic => do
  locExt.set l
  evalTactic rs
  \end{lstlisting}
\end{subfigure}
\captionsetup{aboveskip=2pt}
\captionsetup{belowskip=-10pt}
\caption{Implementing rewrite patterns by passing the rewrite location
  in a local state extension.}
\label{fig:ssrRw}
\end{figure}

To demonstrate the use of this technique, \autoref{fig:ssrRw} shows an
implementation of a restricted version of \lssr rewriting, which only
supports rewrites of the form:
\vspace{-3pt}
\begin{center}
    \code{srw Eq1 // Eq2 Eq3 // ...}
\end{center}
\vspace{-3pt}
In other words, it only allows for rewriting with equality facts (\ie,
\code{Eq1}, \code{Eq2}, \etc.) and for employing last-mile automation
to solve the generated sub-goals.
Importantly, our example also allows for rewriting at a specified
hypothesis \code{H} from the local context using \code{srw ... at H}
syntax.
The syntactic category for rewrite patterns is defined at lines~1-5 of
\autoref{fig:ssrRw}. 
It includes automation (line~2) and ordinary Lean terms (line~3).
Preparing to pass the rewrite location around as an additional state
component, we first define a type of the state extension
\code{LocationExtState} (lines~7-8). It is an option, where \code{some
  ...} denotes a local hypothesis to rewrite at, and \code{none}
stands for the goal.
Next, the \code{initialize} command registers the custom state
component of the type \code{LocationExtState} with the name
\code{locExt}, storing \code{none} to it as a default value
(lines~10-12).
Fast-forward to lines~24-26, the elaboration rule for \code{srw} first
sets the value \code{locExt} to the optional argument \code{l}, \ie,
the hypothesis passed after \code{at} or the goal, if that part is
omitted.
Finally, the elaboration rule for named rewrite patterns (\ie,
equality facts) at lines~19-21 first fetches the location from the
\code{locExt} component of the extended state, and then executes
Lean's native rewrite tactic \code{rw} at that location.

\vspace{-6pt}

\subsection{Computational Reflection via Type Classes}
\label{sec:ssrimpl}

We conclude this section by discussing the implementation of
computational reflection in \lssr via Lean type classes and its
interaction with \code{Decidable} instances.
%

Lines~1-4 of \autoref{fig:ssrimpl} provide the definition of the
\code{Reflect} predicate adopted almost verbatim from \ssr (we will
explain the \code{outParam} keyword below). 
Lean users can notice that \code{Reflect} is very similar to
\code{Decidable} with the only difference being that the former mentions the
Boolean representation of the decidable predicate explicitly.
The \code{#reflect} command at line~10 will use this explicit Boolean
representation~\code{b} to generate reduction/rewrite rules for the
respective predicate \code{P}. This simplification machinery for
propositions is generated from their corresponding symbolic
counterparts in two steps:
\begin{enumerate}[(i),topsep=2pt,leftmargin=17pt]
\item\label{step1} The reduction principles of symbolic representations
  are retrieved as quantified equalities.
\item\label{step2} Boolean functions in those equalities are replaced
  with their corresponding propositions.
\end{enumerate}
Step~\ref{step1} comes ``for free'':
to partially evaluate recursive definitions, Lean automatically
generates lemmas in the form of quantified equalities representing the
available reductions.
For example, for the \code{evenb} definition from \autoref{fig:refl}
in \autoref{sec:refl} Lean will generate tree lemmas:
\begin{center}
  \code{eq1: evenb 0 = true} \quad\quad  
  \code{eq2: evenb 1 = false} \quad\quad  
  \code{eq3: @all n, evenb (n + 2) = evenb n}
\end{center}
Those lemmas are then implicitly used by the \code{simp} and
\code{dsimp} tactics in Lean proofs.

\begin{figure}[t]
  \centering
\begin{subfigure}{0.47\textwidth}
\begin{lstlisting}[basicstyle=\scriptsize\ttfamily,numbers=left,numberstyle=\tiny\color{ACMRed}]
class inductive Reflect (P : Prop) : 
  (b: outParam Bool) -> Type
  | T (_ : P) : Reflect P true
  | F (_ : ¬ P) : Reflect P false

theorem toPropEq (_: b1 = b2) 
  [Reflect P1 b1] [Reflect P2 b2] : 
  P1 = P2 := ...

elab "#reflect" P:term b:term : command => ...
\end{lstlisting}
\end{subfigure}
\hspace{4mm}
\begin{subfigure}{0.50\textwidth}
  \begin{lstlisting}[basicstyle=\scriptsize\ttfamily,firstnumber=11,numbers=left,numberstyle=\tiny\color{ACMRed}]
macro "reflect" : attr => `(attr| default_instance)
@[reflect] instance tP : Reflect True true := ...
@[reflect] instance : Reflect False false := ...
@[reflect] instance [Reflect P1 b1] [Reflect P2 b2]
  : Reflect (P1 ∧ P2) (b1 && b2) := ...
@[reflect] instance [Reflect P1 b1] [Reflect P2 b2]
  : Reflect (P1 ∨ P2) (b1 || b2) := ...
... -- other basic reflect predicates

instance [Reflect P b] : Decidable P := ...
\end{lstlisting}
\end{subfigure}
\captionsetup{aboveskip=2pt}
\captionsetup{belowskip=-15pt}
\caption{A fragment of computational reflection implementation in
  \lssr.}
\label{fig:ssrimpl}
\end{figure}

With all those equalities at hand, step~\ref{step2} is achieved
with the help of the \code{toPropEq} lemma (lines~6-8) that derives the
equalities on propositions out of equalities on their Boolean
counterparts.
For example, applying \code{toPropEq} to the quality \code{eq1} above
will give us the equality \code{even 0 = True}, assuming we also
somehow synthesise its two implicit \code{Reflect} instance arguments.
The synthesis is enabled by the following \lssr components.

The first one is a library of \code{Reflect}-lemmas connecting
standard logic connectives, such as $\wedge$ and $\vee$ with their
computational counterparts, \ie, \code{&&} and \code{||}.
The need for those lemmas arises when we need to synthesise a
propositional version of a symbolic expression that does not have an
application of the function in question at the top level.
In particular, while the left-hand side of all retrieved equalities is
always a function applied to some arguments (\eg, \code{evenb (n +
  2)}), the right-hand-side might be an arbitrary expression
containing other symbolic logical constants and connectives.
The various \code{Reflect}-lemmas (lines~12-18 of
\autoref{fig:ssrimpl}) provide recipes for synthesising \emph{proofs}
for the corresponding logical counterparts in a way similar to
type class instance resolution in Haskell.
\lssr provides a collection of such lemmas, and the user can add their
own, extending the reflection in a modular fashion.

Alas, the type class resolution mechanism of Lean will not immediately
work as desired.
For example, when we apply \code{toPropEq} to \code{eq1}, it will have
to work out two instantiations (\ie, proofs): for \code{Reflect ?P
  (evenb 0)} with \code{?P := even 0} and for \code{Reflect ?Q true}
with \code{?Q := True} (line~12).
By default, Lean's algorithm for synthesising instances will fail due
to a simple reason: to construct an instance of, \eg, \code{Reflect ?P
  (evenb 0)} it \emph{must know} what that \code{?P} is.
That is, the algorithm will only attempt to synthesise an inhabitant
for a fully instantiated type class signature, but not for the one
that has variables.
To fix this, we use the mechanism of \emph{default instances}, \ie,
proof terms that are available for instance search even when not
all type class arguments are known~\cite[\S{4.3}]{funlean}.
For example, marking \code{tP : Reflect True true} as a default
instance will make the type class resolution algorithm aware of it,
thus, finding a suitable instantiation for \code{?Q} in the process.
For readability, \lssr provides a \code{reflect} notation for the
\code{default_instance} attribute, so that \code{Reflect} instances
can be marked as default ones simply by tagging them with \code{@[reflect]}.

The last bit of our implementation is a link between \code{Reflect}
and \code{Decidable}.
Once a \code{Reflect} instance for a logical predicate \code{P} is
provided, \lssr also generates its decidable instance (line~20 of
\autoref{fig:ssrimpl}), \eg, to use this \code{P} in any position that
requires a Boolean expression.
Marking the parameter~\code{b} of \code{Reflect} as \code{outParam}
(line~2) is essential for this: it tells the instance search algorithm
that \code{b} is not required for finding the instance (since \code{P}
already provides enough information), and should be treated as the
output of the synthesis.



\section{\lssr in the Wild: Case Studies}
\label{sec:case}


In this section, we showcase two examples that support our claims from
\autoref{sec:intro}, by demonstrating \lssr's expressivity compared to
Coq/\ssr and to vanilla Lean, highlighting the more compact proof
scripts we obtain, as well as the overall usability of our approach.


\subsection{Migrating MathComp Sequences to \lssr}
\label{sec:ssrseq}

To evaluate \lssr against \ssr, we ported a small subset (approximately 10\%) of
the finite sequences file from the Mathematical Components
library Coq~\cite{Maboubi-Tassi:MathComp}, amounting to 31 definitions, 72 theorems,
and 3 reflection predicates.
As the logics of Lean and Coq are similar, and MathComp is implemented in \ssr,
which is syntactically very similar to \lssr, most proofs can be translated
almost mechanically, with minimal changes.

\begin{figure}[t]
\begin{lstlisting}[language=SSR,basicstyle=\footnotesize\ttfamily,numbers=left,numberstyle=\tiny\color{ACMRed},
  moredelim={**[is][{\btHL[fill=orange!30]}]{@}{@}},
  moredelim={**[is][{\btHL[fill=red!30]}]{`}{`}},]
Lemma subseq_trans: transitive subseq.
Proof.
move=> _ _ s `/subseqP` [m2 _ ->] `/subseqP` [m1 _ ->].
elim: s m1 m2 => [*|x s IHs]; @first by rewrite !mask0.@
case=> [*|[] m1 m2]; @first by rewrite !mask0.@
{ case: m2=> [/=|[] m2] //; @first by rewrite /= eqxx IHs.@
  case `/subseqP`: (IHs m1 m2) => m sz_m ?; `apply/subseqP.`
  by exists (false :: m); @rewrite //= sz_m.@ }
case`/subseqP`: (IHs m1 m2)=> m sz_m ?; `apply/subseqP.`
by exists (false:: m); @rewrite //= sz_m.@
Qed.
\end{lstlisting}
\captionsetup{aboveskip=0pt}
\captionsetup{belowskip=-10pt}
\caption{\ssr proof of the theorem in \autoref{fig:trans}. Parts not needed in \lssr are highlighted.}
\label{fig:subseqCoq}
\end{figure}

Our proofs are nonetheless more compact than the originals for two reasons:
(a) the user does not need to explicitly switch between logical and symbolic
representations of the goal (claim~\ref{label:express})
and (b) many rewrites can be performed by \lssr's simplification mechanism, due
to its extensibility (claim~\ref{label:ext}), and do not need to be invoked
manually.
A representative example can be seen in \autoref{fig:subseqCoq}, which shows a
\ssr proof (slightly modified from the original for presentation purposes) of
the fact that the subsequence relation is transitive.
The \lssr equivalent was shown previously in \autoref{fig:trans}.
The Lean version is more compact as it requires neither the explicit invocations
of the reflection predicate (highlighted in \redbox{red} in
\autoref{fig:subseqCoq}) nor various trivial rewrites (highlighted in
\orangebox{orange}), which can be performed automatically by \lssr's powerful
and extensible \code{//} automation.
Overall, this amounts to a reduction in half of the size of each line of proof,
and we argue, to a reduction of the cognitive effort required of the user, who
can now focus on the essential aspects of the proof, delegating trivial details
to automation.
Indeed, many simple statements can be proven in \lssr by \code{elim:
  s=>//=}, reminiscent of the \code{induct-and-simplify} tactic of
PVS~\cite{owre1992pvs, shankar2001pvs}.


\subsection{Refactoring mathlib}
\label{sec:mathlib}

To evaluate \lssr against proofs in mathlib~\cite{mathlib20},
we ported a few facts about cardinalities of finite sets.
\autoref{fig:lssr-vs-mathlib} presents two proofs of \emph{a subgoal} for
the \code{card\_eq\_of\_bijective} lemma from mathlib: the original proof and
one ported to \lssr. The subgoal states that for a bijective
function~\code{f} from a range of all natural number less
than~\code{n}, each element in this range has a pre-image
\wrt~\code{f}.
Notably, mathlib proofs frequently build a proof term explicitly,
rather than via tactics.
In cases like the one in \autoref{fig:lssr-vs-mathlib}, such
manual proof term construction tends to be more concise than the
respective proof script, hence the ubiquity of this mechanisation
style in mathlib.
%
%
The same proof ported to \lssr is shown in \autoref{fig:lssr-mathlib}.
Here, the \code{@langle .. | ... | .. ⟩} pattern is similar to \code{[ .. | ...
  | .. ]}, but instead of matching on the top element of the goal
stack, it applies the \code{constructor} tactic to the goal, and runs
nested tactics on the generated subgoals.
The main difference between those two proofs is that instead of the
somewhat awkward backward-style reasoning with explicitly constructed
proof terms (\eg, line 10 in \autoref{fig:mathlib}), \lssr allows for natural
forward-style proofs using \lssr views (\cf \autoref{sec:views}).
%
%
%
%
Specifically, the mathlib proof of the second component of the
bi-implication from the conclusion (\ie, right-to-left) first
destructs the hypothesis \code{@ex i, ∃ (hi : i ∈ range n), f i _ = a},
and then gradually reduces the goal (\code{a ∈ s}) to the assumption
\code{hi : i ∈ range n} by first rewriting \code{eq}, then applying
\code{hf'} and finally \code{mem_range : n ∈ range m ↔ n < m}.
In contrast,the  \lssr proof adapts \code{hi} to solve the goal
automatically at line~8 of \autoref{fig:lssr-mathlib}, by first
applying \code{mem_range} to its second existential, then \code{hf'}
to the result, and then rewriting \code{eq} in it via the special form
of the view pattern \code{/[swap]}, followed by \code{->}.

\begin{figure}[t]
  \centering
\begin{subfigure}{0.51\textwidth}
\begin{lstlisting}[basicstyle=\scriptsize\ttfamily,numbers=left,numberstyle=\tiny\color{ACMRed}]
example     
  (hf : ∀ a ∈ s, ∃ i (h : i < n), f i h = a)
  (hf' : ∀ i (h : i < n), f i h ∈ s) :
  ∀ (a : α), a ∈ s ↔ 
    ∃ i, ∃ (hi : i ∈ range n), f i _ = a := 
  fun a =>
    ⟨fun ha =>
      let ⟨i, hi, eq⟩ := hf a ha
      ⟨i, mem_range.2 hi, eq⟩,
      fun ⟨i, hi, eq⟩ => eq ▸ hf' i (mem_range.1 hi)⟩
\end{lstlisting}
\caption{Original mathlib proof script}
\label{fig:mathlib}
\end{subfigure}
\hspace{4mm}
\begin{subfigure}{0.44\textwidth}
  \begin{lstlisting}[basicstyle=\scriptsize\ttfamily,numbers=left,numberstyle=\tiny\color{ACMRed}]
example     
  (hf: ∀ a ∈ s, ∃ i (h : i < n), f i h = a)
  (hf' : ∀ i (h : i < n), f i h ∈ s) :
  ∀ (a : α), a ∈ s ↔ 
    ∃ i, ∃ (hi : i ∈ range n), f i _ = a :=
  by move=> a
    ⟨/hf ![i /mem_range ? <-] // | 
    ![i /mem_range /hf' /[swap]->] //⟩
\end{lstlisting}
\vspace{16pt}
\caption{\lssr proof script}
\label{fig:lssr-mathlib}
\end{subfigure}
\caption{Two proofs of the same fact from mathlib: the vanilla one and
  the one in \lssr.}
\label{fig:lssr-vs-mathlib}
\end{figure}

We claim that the presented \lssr proof has certain advantages compared to
the one from mathlib.
First, views will automatically infer dependent arguments when
applying a lemma, \eg, \code{i} in \code{hf' i mem_range.1 hi} at
line~10 of \autoref{fig:mathlib}.
Second, \lssr organically incorporates last-mile automation (using
\code{//} and other automation patterns) to the forward-style proof
script, thus, sparing the user the effort of building yet another
collection of proof terms to dispatch the subgoals.
Finally, unlike the proof terms constructed holistically, \lssr proofs
are interactive: one can check the intermediate subgoals by simply
pointing to the specific location in the text buffer, resulting in a
better overall proof experience.
In summary, our experiment demonstrates that \lssr proofs are not only shorter
compared to the original mathlib ones, but also easier to follow
and debug.
The latter aspect is crucial for the long-term maintainability of large
formal Lean developments such as mathlib.

\section{Related Work and Conclusion}





Prior to our effort, Lean 4's metaprogramming facilities have been
used to implement white-box automation via proof search in the popular
Aesop package~\cite{aesop23}.
In particular, mathlib~\cite{mathlib20} uses Aesop to build
domain-specific solvers such as \code{measurability} and
\code{continuity}, and uses metaprogramming extensively to define its
own tactics and automation facilities (\eg, \code{split_ifs} and
\code{linarith}).
Closer in spirit to our work, K\"{o}nig developed a proof interface in
Lean in the stlye of Iris proof mode~\cite{KrebbersJ0TKTCD18},
featuring a specialised tactic language for proofs in Separation
Logic~\cite{konig22}.
In Lean~3, Limperg built an induction tactic that is friendlier for
novices by giving the most general induction
hypothesis~\cite{Limperg21}.

We believe that our implementation of \lssr provides an instructive
example of using Lean~4 metaprogramming features for implementing a
non-trivial tactic language, and adds one more arrow to the quiver of
tools and techniques for proof construction in Lean.


\bibliography{references}

\end{document}